\documentclass[conference]{scrartcl}
\usepackage[ansinew]{inputenc}
\usepackage[T1]{fontenc}
\usepackage[english]{babel}
\usepackage{color,multirow}
\usepackage{amsmath, amssymb, amsthm}
\usepackage[Gray,squaren,thinqspace,thinspace]{SIunits}
\usepackage{xcolor} 
\usepackage{flafter}
\usepackage{float}
\usepackage{subcaption}
\usepackage{lipsum}
\usepackage{booktabs}
\usepackage{xcolor}
\usepackage{graphicx}
\usepackage{caption}
\usepackage{subcaption}
\usepackage[affil-it]{authblk}

\usepackage{eso-pic}

\AddToShipoutPicture*{\footnotesize\sffamily\raisebox{2.5cm}{\hspace{3cm}\fbox{\parbox{\textwidth}{\copyright~2017
IEEE. Personal use of this material is permitted. Permission from IEEE
must be obtained for all other uses, in any current or future media,
including reprinting/republishing this material for advertising or
promotional purposes, creating new collective works, for resale or
redistribution to servers or lists, or reuse of any copyrighted
component of this work in other works.}}}}

\newcommand{\mean}{\mathbb{E}} 

\newcommand{\var}{\mathbb{V}} 
\newcommand{\randV}{\theta}
\newcommand{\Jsource}{J_{\mathrm{s}}}

\newcommand{\order}{\mathcal{O}}
\newcommand{\MSE}{\textrm{MSE}}

\begin{document}

\title{Multilevel Monte Carlo Simulation of the Eddy Current Problem With Random Parameters}

\date{}
\author[1]{\large Armin Galetzka}
\author[2]{Zeger Bontinck}
\author[1,2]{Ulrich R\"omer}
\author[1,2]{Sebastian Sch\"ops}
\affil[1]{Institut f\"ur Theorie Elektromagnetischer Felder, ~\newline Technische Universit\"at Darmstadt, Darmstadt, Germany ~\newline
Email: armin\_herbert.galetzka@stud.tu-darmstadt.de, roemer@temf.tu-darmstadt.de \\}
\affil[2]{Graduate School of Computational Engineering~\newline Technische Universit\"at Darmstadt, Darmstadt, Germany~\newline
Email: $\{$bontinck, schoeps$\}$@gsc.tu-darmstadt.de}

\maketitle

\begin{abstract}
The multilevel Monte Carlo method is applied to an academic example in the field of electromagnetism.
The method exhibits a reduced variance by assigning the samples to multiple models with a varying spatial resolution. For the given example it is found that the main costs of the method are spent on the coarsest level.
\end{abstract}

\section{Introduction}
The two most known approaches to propagate uncertainties in the field of uncertainty quantification rely on generalized Polynomial Chaos (gPC) expansions \cite{Xiu_2002aa} or use the classical Monte Carlo (MC) method. When the number of random parameters is small and the solution depends in a smooth way on these parameters, a gPC approach is known to have a superior convergence rate \cite{Xiu_2002aa}. However, as the number of uncertain parameters $d$ increases, the method becomes inefficient. This is often called the curse of dimensionality (see e.g.~\cite{Kuo_2005aa}). For a large number of random variables and a lack of regularity one has to rely on MC methods, since their convergence doesn't depend on $d$. However, the error of classical MC converges slowly with $\order(N^{-1/2})$, with $N$ the necessary number of samples. 
\begin{figure}[t] 
  \centering
  \includegraphics[width=0.26\textwidth]{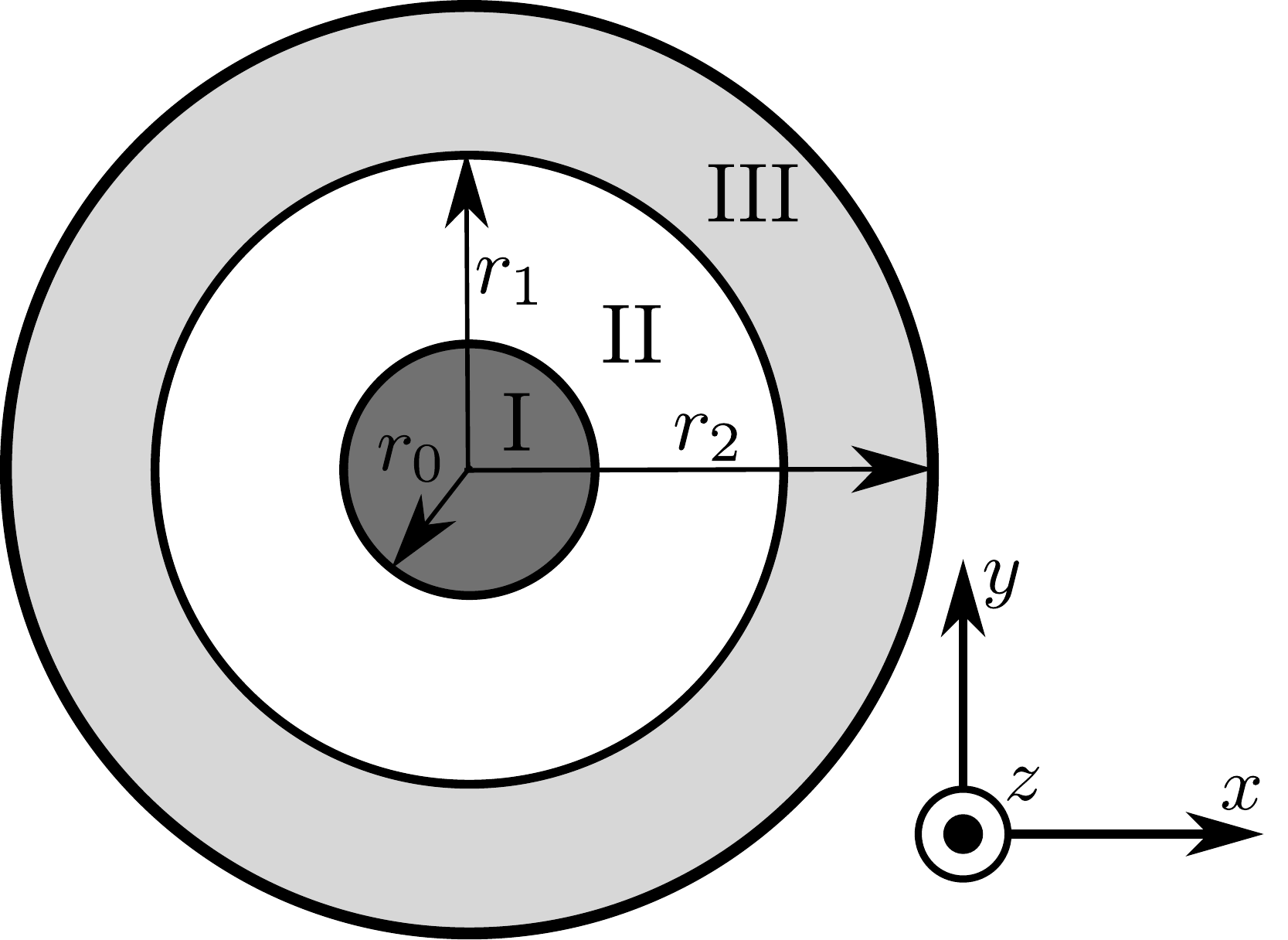}
  \caption{Cross-sectional view of the conducting wire (region I) in a steel tube (region III). Region II is filled with air.}
  \label{fig:tube}\vspace*{-0.4em}
\end{figure}
Many efforts have been undertaken to overcome the slow convergence rate. One way is to use quasi-MC methods e.g.~\cite{Dick_2013aa}. Another way out is using multilevel Monte Carlo (MLMC) \cite{Giles_2015aa}. In this paper every level refers to a spatial discretization level. The gain in the cost of MLMC with respect to MC originates from the fact that a big part of the uncertainty can be captured by using models with coarse spatial discretization. This means that the number of evaluations on the finest grid can be reduced drastically. The idea of this paper is to apply the MLMC method to an academic example in the field of electromagnetism. Also a study of the computational costs is done. More details on MLMC in the context of (elliptic) partial differential equations, can be found in \cite{Teckentrup_2013aa}.

\section{Application}
Let us consider a conducting wire centered in a steel tube, Fig.~\ref{fig:tube}. The region between the wire and the tube is filled with air. The steel pipe has a relative permeability $\mu_{\mathrm{III}}$ and conductivity $\sigma$. The wire and the air region are modeled by vacuum permeability $\mu_0=\mu_{\mathrm{I}}=\mu_{\mathrm{II}}$ but without conductivity. The wire has a source current density $\Jsource=I_0\sin(\omega t)/\pi r^2_0$, with $I_0$ the magnitude of the current, $r_0$ the radius of the wire and $\omega$ the angular frequency.
As a test case we consider three ($d=3$) random variables: $r_1$, which is the inner radius of the steel tube, $\mu_{\mathrm{III}}$ and $I_0$. They are uniformly distributed
\begin{alignat}{2}
r_1(\theta)&=\bar{r}_1+X(\randV), \,\,\,\,\,\,\,\,\,\,\,\,\,\,&&X(\randV)\sim\mathcal{U}(-0.1\,\textrm{m}, 0.1\,\textrm{m}), \label{eq:boundaryR1}\\
I_0(\theta)&=\bar{I}_0+Y(\randV), &&Y(\randV)\sim\mathcal{U}(-10\,\textrm{A}, 10 \,\textrm{A}), \\
\mu_{\mathrm{III}}(\theta)&=\bar{\mu}_{\mathrm{III}}+Z(\randV), &&Z(\randV)\sim\mathcal{U}(-400, 400),
\end{alignat}
where $\randV$ depicts the random outcome of a quantity. The nominal values are defined by
\begin{alignat}{2}
\bar{r}_1 &= \mean[r_1(\randV)] &&= 0.5 \, \textrm{m}, \\
\bar{I}_0 &= \mean[I_0(\randV)] &&= 100 \, \textrm{A}, \\
\bar{\mu}_{\mathrm{III}} &= \mean[\mu_{\mathrm{III}}(\randV)] &&= 1000,
\end{alignat}
where $\mean[\cdot]$ denotes the expectation value. The problem can be treated by the magnetoquasistatic approximation of Maxwell's equations.
\begin{figure*}[t!]
  \vspace*{-0.4em}
  \centering
  \begin{subfigure}{0.48\textwidth}
    \centering
       \includegraphics[scale=0.5]{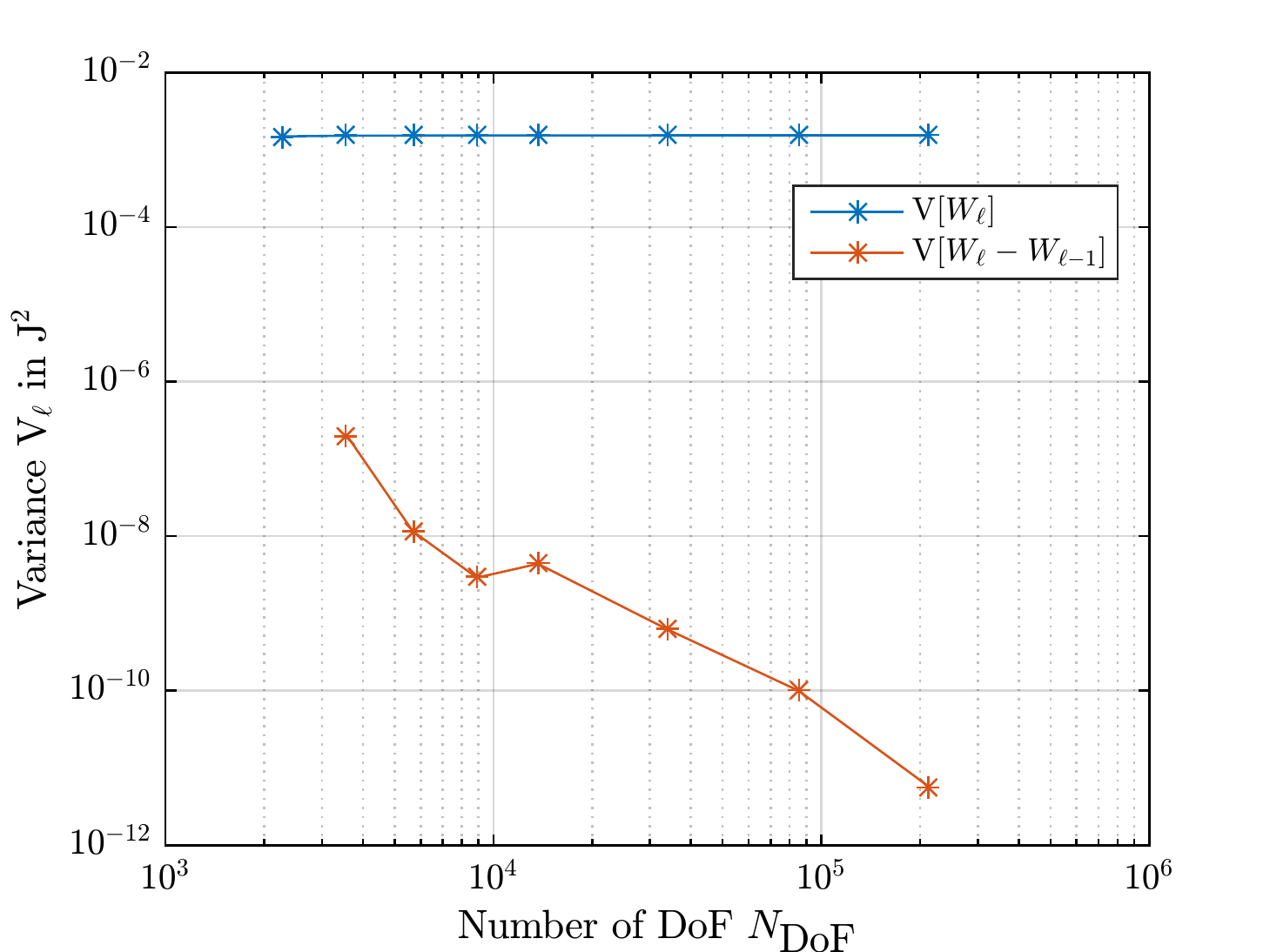}
             \caption{$\var[W_\ell]$ and $\var[W_\ell-W_{\ell-1}]$ over $N_{\mathrm{DoF}}$.}
        \label{fig:varianceGPCFEM}
  \end{subfigure}
  \hspace{0.01\textwidth}
  \begin{subfigure}{0.48\textwidth}
    \centering
     \vspace*{0.5em}
       \includegraphics[scale=0.54]{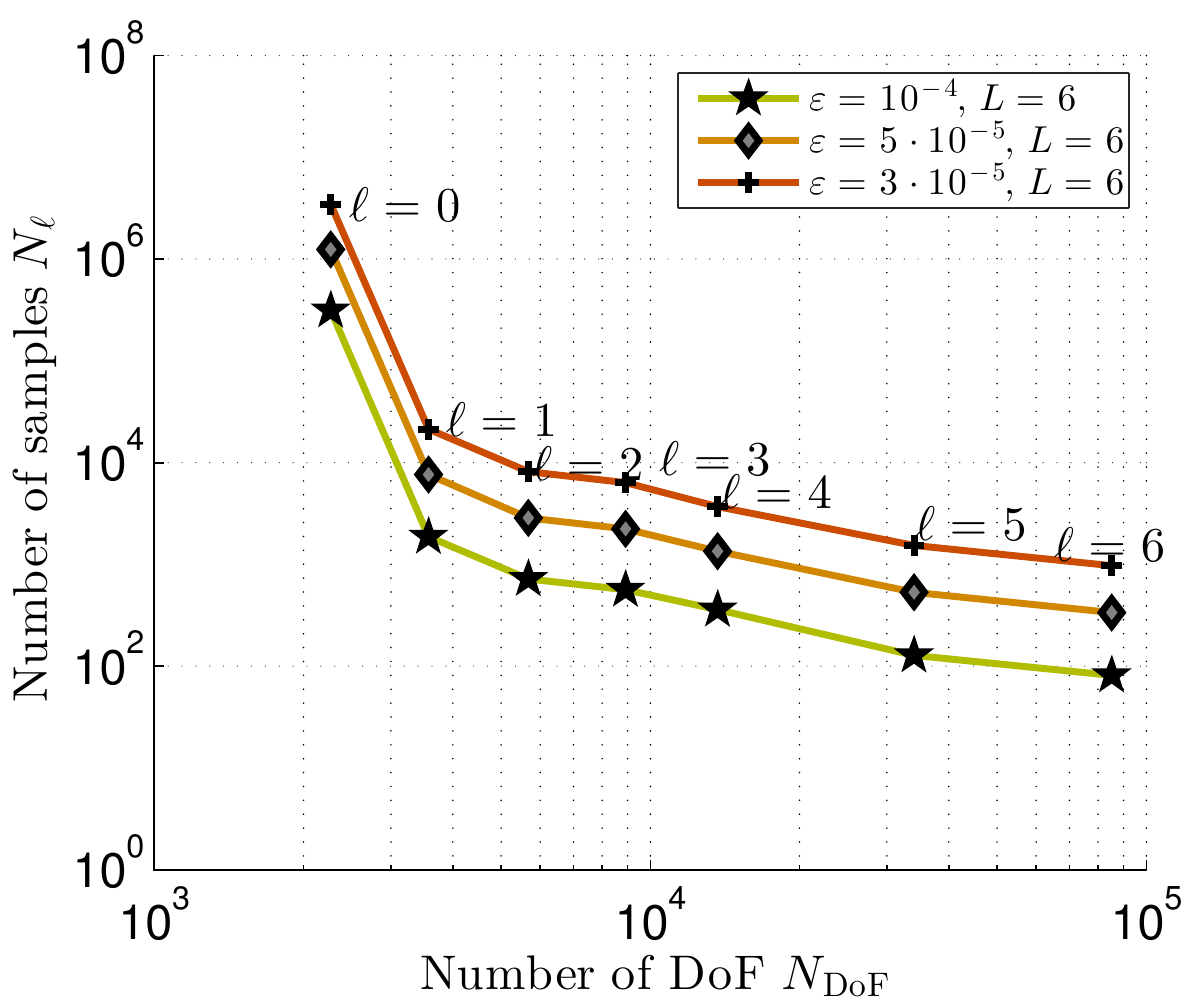}
             \caption{Number of samples $N_\ell$ over $N_{\mathrm{DoF}}$.}
        \label{fig:samplesLvlGama3}
  \end{subfigure}
  \hspace{0.01\textwidth}
  \caption{Variance and number of samples in dependence of DoFs.}\vspace*{-0.5em}
\end{figure*}
Modeling the source current density as a time harmonic function one retrieves 
\begin{equation}
\nabla \times \left( \frac{1}{\mu(\randV)} \nabla \times \vec{A}(\randV) \right) + j \omega \sigma \vec{A}(\randV) = \vec{J}_s(\randV),
\label{eq:SPDE_FRQ}
\end{equation}
where $\vec{A}$ is the magnetic vector potential. Imposing Dirichlet boundary conditions on the outer radius of the pipe one can solve~\eqref{eq:SPDE_FRQ} for $\vec{A}$. This allows the calculation of the energy
\begin{equation}
W(\randV) = \sum_{i \in \{\textrm{I},\textrm{II},\textrm{III}\}} \int_{V^{(i)}}\frac{\mu_{i}\vec{H}^{(i)}(\randV)\cdot \underline{\vec{H}}^{(i)}(\randV)}{2}\textrm{d}V,
\label{eq:magneticEnergyGeneral}
\end{equation}
where $\vec{H}^{(i)}(\randV)=\frac{1}{\mu_i(\randV)}\nabla\times\vec{A}^{(i)}(\randV)$ is the magnetic field and $i\in\{\mathrm{I},\mathrm{II},\mathrm{III}\}$ refers to the regions.

Since it assumed that $\vec{A}=(0,0,A_z)$,~\eqref{eq:SPDE_FRQ} is solved in 2D and lowest order finite elements are used. The discretization is done by a triangular grid~\cite{Salon_1995aa}.

\section{Multilevel Monte Carlo}
The main principles of MLMC are recalled from \cite{Giles_2015aa} in the following.
Let $\mean[W]$ be the quantity of interest and $W_\ell$ be the energy on level $\ell=0,\dots,L$, with 0 and $L$ referring to the coarsest and finest level, respectively. Also, let $Y$ denote an approximation to $\mean[W]$, i.e. the MC or MLMC estimator. For the mean square error (MSE) there holds
\begin{align}
\MSE 
	= \var[Y] + \left(\mean[Y-W]\right)^2.
	\label{eq:meanSquareError}
\end{align}
Hence, an overall MSE of $\varepsilon^2$ can be achieved, by reducing both the weak error $\left(\mean[Y-W]\right)^2$ and the variance $\var[Y]$ below $\varepsilon^2/2$. In MLMC, the weak error is dominated by the spatial resolution of the finest level and can be controlled accordingly.

\begin{figure}[t!]
  \centering 
       \includegraphics[scale=0.6]{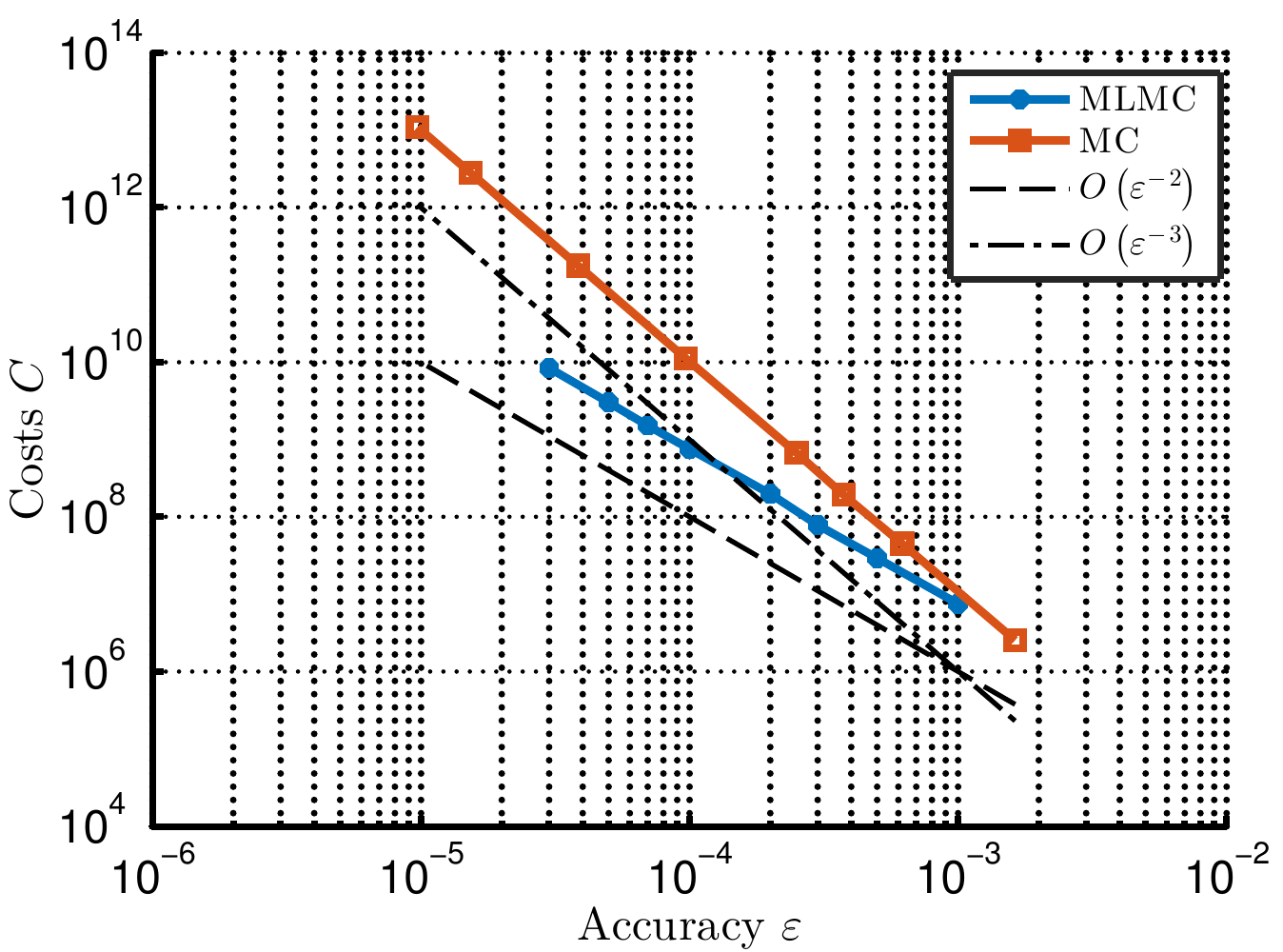}
       \caption{$Y$ evaluated for different error bounds $\varepsilon$ over computational costs.}
       \label{fig:CostsMC_MLMC_Gamma3}
  \end{figure}

The variance is reduced by adding more and more samples to the different levels. More precisely, by decomposing $Y$ as
\begin{equation}
\label{eq:Y}
Y=\mean[W_0] + \sum_{\ell=1}^L \mean\left[W_\ell-W_{\ell-1}\right],
\end{equation}
the conventional MC method is used with $N_\ell$ samples for each term on the rhs in \eqref{eq:Y}. The associated variance reads
\begin{equation}
\label{eq:var}
\var[Y]=N_0^{-1}\var[W_0]+\sum_{\ell=1}^L N_\ell^{-1}\var[W_\ell - W_{\ell-1}].
\end{equation}
For a fixed computational budget, there exists an optimal choice of $N_\ell$ minimizing the variance given by \eqref{eq:var}, see \cite{Giles_2015aa}. This optimum is readily computable if asymptotic bounds on $\var[W_\ell - W_{\ell-1}]$ and the cost per level can be determined. The coefficients in these bounds are problem dependent and determined numerically in this work. 

\section{Results}
The variance $\var[W_\ell]$ is nearly constant over all levels as shown in Fig.~\ref{fig:varianceGPCFEM} and thus independent of the number of degrees of freedom (DoF) $N_{\mathrm{DoF}}$. This observation confirms the fact that the MSE can be divided into an error determined by the variance of the estimator and an error determined by the FEM approximation. In Fig.~\ref{fig:samplesLvlGama3} the number of samples is plotted. Different error bounds are used. As anticipated, the lower the bounds the more samples are needed. Taking more samples also implies an increase in the overall computational cost for every level. It is found that the biggest computational cost is spent on the coarsest level. The reason can be seen by looking to Fig.~\ref{fig:varianceGPCFEM}. More samples are needed on the coarsest level for the convergence of the variance, however less samples are needed on the finer levels since $\var[W_\ell-W_{\ell-1}]$ is small. In Fig.~\ref{fig:CostsMC_MLMC_Gamma3} the costs $C=N_{\mathrm{DoF}}N_{\mathrm{MC}}$ of MLMC and MC are compared in dependence of the demanded accuracy. The number of MC samples is depicted by $N_{\mathrm{MC}}$. One clearly sees that for $\varepsilon<10^{-3}$ the costs of MLMC are lower than the ones using~MC.

\section{Conclusion}
The cost of MLMC is determined by the costs on the coarsest level since the reduction of the variance is dominating. Adding more levels does not alter the variance much, but helps reducing the weak error. Comparing the costs of MLMC with classical MC, one sees that for an accuracy $\varepsilon<10^{-3}$ the costs of MLMC are lower than the ones of MC.

\section*{Acknowledgment}
This work is supported by the German BMBF in the context of the SIMUROM project, by the 'Excellence Initiative' of the German Federal and State Governments and the Graduate School of CE at TU Darmstadt.

\end{document}